# BaFe$_2$(As$_{1-x}$P$_x$)$_2$ ($x$ = 0.22–0.42) thin films grown on practical metal-tape substrates and their critical current densities

Hidenori Hiramatsu[1,2,*], Hikaru Sato[1], Toshio Kamiya[1,2], and Hideo Hosono[1,2]

[1] Laboratory for Materials and Structures, Institute of Innovative Research, Tokyo Institute of Technology, Mailbox R3-1, 4259 Nagatsuta-cho, Midori-ku, Yokohama 226-8503, Japan

[2] Materials Research Center for Element Strategy, Tokyo Institute of Technology, Mailbox SE-6, 4259 Nagatsuta-cho, Midori-ku, Yokohama 226-8503, Japan

*E-mail: h-hirama@lucid.msl.titech.ac.jp

Abstract

We optimized the substrate temperature ($T_s$) and phosphorus concentration ($x$) of BaFe$_2$(As$_{1-x}$P$_x$)$_2$ films on practical metal-tape substrates for pulsed laser deposition from the viewpoints of crystallinity, superconductor critical temperature ($T_c$), and critical current density ($J_c$). It was found that the optimum $T_s$ and $x$ values are 1050 °C and $x = 0.28$, respectively. The optimized film exhibits $T_c^{onset} = 26.6$ and $T_c^{zero} = 22.4$ K along with a high self-field $J_c$ at 4 K (~1 MA/cm$^2$) and relatively isotropic $J_c$ under magnetic fields up to 9 T. Unexpectedly, we found that lower crystallinity samples, which were grown at a higher $T_s$ of 1250 °C than the optimized $T_s = 1050$ °C, exhibit higher $J_c$ along the *ab* plane under high magnetic fields than the optimized samples. The presence of horizontal defects that act as strong vortex pinning centers, such as stacking faults, are a possible origin of the high $J_c$ values in the poor crystallinity samples.

1. Introduction

High critical temperature ($T_c$) Fe-based superconductors were discovered in 2008 [1]. Superconductor researchers were initially interested in an unpredictability that a ferromagnetic Fe-containing compound exhibits a high $T_c$ of 26 K, and they then started exploring related materials for new high-$T_c$ superconductors [2,3]. With extensive research, it has been clarified that Fe-based superconductors possess suitable properties for superconducting wire/tape/coated conductor applications, such as high-field magnets and electric power cables, including high upper critical magnetic fields (>>50 T) [4], low anisotropy factors (~1–7) [5], and high $T_c$ values (maximum ~55 K) [6]. Therefore, many superconducting wire/tape [7–10] and coated conductors [11–19] have been developed using Fe-based superconductors and their performance has been rapidly improving. Another important characteristic of Fe-based superconductors for application to wire/tape/coated conductors is the advantageous grain boundary nature. For example, the iron-based superconductor $Ba(Fe_{1-x}Co_x)_2As_2$ (Ba122:Co) has a considerably higher critical grain-boundary misorientation angle ($\theta_c$ = ~9°) for the critical current density ($J_c$) than those of high-$T_c$ cuprates ($\theta_c$ = ~4°) [20]. This advantage over cuprates enables lower cost fabrication processes for superconducting wire/tape/coated conductors using less aligned grain orientations. Thus, low-cost technical metal-tape substrates can be used for Fe-based superconductors.

In thin-film research of Fe-based superconductors, Ba122:Co films have been the most extensively investigated because of the easiness of performing high-quality film growth, mainly owing to the low vapor pressure of the Co dopant [21]. Because $BaFe_2(As_{1-x}P_x)_2$ (Ba122:P) exhibits a higher maximum $T_c$ (~31 K) [22] than Ba122:Co (~22 K) [23], Ba122:P films have also been extensively investigated [24–29]. From the viewpoint of application to coated conductors, we recently investigated Ba122:P films on practical metal-tape substrates [17]. These films were produced based on biaxially textured MgO buffer layers with top epitaxial MgO layers on flexible non-oriented metal-tapes grown by the ion beam assisted deposition (IBAD) method. Unexpectedly, the $J_c$ of Ba122:P films on "poorly in-plane aligned" IBAD-MgO metal-tape substrates, where the Ba122:P films have much poor in-plane crystallinity, was higher than that on "well in-plane aligned" substrates. This apparent contradictory behavior would reflect the above grain boundary nature [20] and be appropriate for future practical applications of Ba122:P coated conductors because a thinner top-MgO epitaxial layer can be used for poorly aligned IBAD-MgO substrates, leading to shorter time and lower-cost fabrication processes. Indeed, recently a poorly aligned Ba122:P coated conductor grown at relatively high temperature exhibits practical $J_c$ performance (0.1 MA/cm$^2$ at 15 T) [18].

In this study, we optimized the film growth conditions for Ba122:P coated conductors on practical IBAD-MgO metal-tape substrates and investigated the electron transport properties, such as $T_c$ and $J_c$, of Ba122:P films with different phosphorus doping concentrations.

2. Experimental details

2.1. Thin film growth

The IBAD-MgO substrates were supplied by iBeam Materials, Inc. (Santa Fe, NM, USA) [30, 31]. The IBAD-MgO substrate consists of a top MgO epitaxial layer, a biaxially textured MgO layer grown by the IBAD method, and a planarizing amorphous Y$_2$O$_3$ bed-layer on a non-oriented polycrystalline Ni-based Hastelloy flexible metal-tape substrate. Ba122:P thin films with 150–200 nm in thickness were grown on the IBAD-MgO substrates (in-plane alignment of the top MgO layer of $\Delta\phi_{MgO} \approx 4°$, lateral size ~10 mm × 10 mm) by pulsed laser deposition (PLD). Polycrystalline $BaFe_2(As_{1-x}P_x)_2$ disks (nominal $x$ = 0.3, 0.35, 0.4, and 0.45) as PLD target disks were synthesized by solid-state reactions of stoichiometric mixtures of the precursor powders (BaAs, $Fe_2As$, and $Fe_2P$) by the reaction $BaAs + (1-x)Fe_2As + xFe_2P \rightarrow BaFe_2(As_{1-x}P_x)_2$ at 930 °C for 16 h in Ar-filled stainless tubes. The substrate temperature ($T_s$) was varied from 800 to 1250 °C. The other growth conditions (e.g., PLD growth chamber, excitation laser source, growth rate, and so on) were the same as the optimized conditions for Ba122:P film growth on MgO single-crystal substrates [28].

2.2. Characterization

$\theta$-coupled $2\theta$-scan X-ray diffraction (XRD) measurements with the Bragg–Brentano geometry were performed to determine the crystalline phases, crystallite orientation, and $c$-axis lattice parameters. The in-plane heteroepitaxy and $a$-axis lattice parameters were confirmed by in-plane and asymmetric XRD measurements [17]. The crystallinity of the films was characterized based on the full width at half maximum (FWHM) of the out-of-plane rocking curves ($\Delta\omega$) of the Ba122:P 004 diffraction. The phosphorous concentration in the films (i.e., $x$ = P/(As+P)) was determined with an electron-probe microanalyzer (EPMA) using the ZAF correction method. All of the characterizations were performed at room temperature.

To investigate the electron transport properties, the films were patterned into microbridges (length 300 μm and width 10 or 20 μm) by photolithography and Ar ion milling. The temperature dependence of the electrical resistivity ($\rho$) was measured by the four-probe method with a physical property measurement system. The transport $J_c$ values at 4 and 12 K were determined from voltage–current curves with the criterion of 1 μV/cm under external magnetic fields ($H$) up to $\mu_0 H$ = 9 T in the maximum Lorentz force configuration ($J \perp H$). The angle of the applied $H$ ($\theta_H$) was varied from −30 to 120°, where $\theta_H$ = 0 and 90° correspond to the configurations of $H \parallel c$ axis of the Ba122:P film (i.e., normal to the film surface) and $H \parallel ab$ plane, respectively.

3. Results and discussion

3.1. Optimization of the film growth conditions

We optimized $T_s$ using a PLD target with nominal $x$ = 0.3. Figure 1(a) shows the $T_s$ dependences of the $c$-axis length and $x$ in the film determined by the EPMA. There are three types of crystallographic orientation structure: (i) random (almost no preferentially orientation), (ii) epitaxial (heteroepitaxially grown films on IBAD-MgO), and (iii) 00$l$ & $hh$0 (mixture of $c$-axis and $ab$-plane orientations). The typical XRD patterns for each type are shown in Fig. 1(b). With increasing $T_s$, the crystallographic orientation changes from random to epitaxial to 00$l$ & $hh$0. A similar variation of the orientation structure is also observed in the $T_s$ dependence of Ba122:Co film growth [32], although its $T_s$ range is lower than that of the present Ba122:P film. At $T_s \approx$ 1000 °C, the amorphous $Y_2O_3$ bed-layers in the IBAD-MgO substrates start to crystallize, as indicated by the inverted triangles in Fig. 1(b), because of the high-$T_s$ growth. The $x$ values in the films are lower than that in the PLD target (i.e., nominal $x$ = 0.3) and slightly increase with increasing $T_s$, which is consistent with the decrease in the $c$-axis length (Fig. 1(a)). The narrowest FWHM of $\Delta\omega$ (1.2°) is achieved at $T_s \approx$ 1050–1100 °C (Fig. 1(c)), which corresponds to the epitaxial region. For $T_s \geq 1200$ °C, the FWHM of $\Delta\omega$ alternates between 1° and 3° at $T_s = 1200$ °C, and exceeds 3° at $T_s = 1250$ °C (Fig. 1(c)). In Ref. [18], the FWHM of Ba122:P grown at 1200 °C on IBAD-MgO with $\Delta\phi_{MgO} = 8°$ is 1.2°. Thus, $T_s = 1200$ °C is the critical point for the out-of-plane crystallinity. From these results, we conclude that the optimum $T_s$ is 1050 °C, which is the same as the case of Ba122:P growth on MgO single-crystals [28], although the $Y_2O_3$ bed-layer in IBAD-MgO slightly crystallizes. $T_s = 1050$ °C is optimum from the viewpoint of the crystallinity, but the films grown at higher $T_s$ (e.g., 1250 °C) exhibit better $J_c$ performance under an external magnetic field, which will be discussed later. The Ba122:P film on IBAD-MgO grown at the optimum $T_s$ has a more relaxed structure than the film on the MgO single-crystal because the *a*- and *c*-axes are both closer to those of the single crystal with the same *x* [22] (see Fig. 1(d)).

We then grew Ba122:P films using PLD targets with different nominal *x* values at the optimum $T_s = 1050$ °C. All of the *x* values in the films are slightly lower than the nominal *x* values, but there is a linear relationship between the measured *x* and the nominal *x* (Fig. 1(e)), indicating that the *x* value in the film can be controlled by the nominal *x* value in the PLD target. As shown in Fig. 1(d), the *c*-axis length of the film on IBAD-MgO is shorter than that of the single crystal but much closer to that of the film on a single-crystal MgO substrate, showing that the crystallites in the film on IBAD-MgO are relaxed and closer to the single-crystal structure. Similarly, the *c*-axis lengths in the films are shorter than those of the single crystals but there is a similar linear *c*–*x* relationship (Fig. 1(f)).

3.2. Electron transport properties

Figure 2 shows the temperature dependence of the normalized $\rho$ for four Ba122:P films with $x$ = 0.22–0.42 grown at the optimum $T_s$ = 1050 °C. The straight lines and $\alpha$ values in Fig. 2(a) are the fitting results using the power law $\rho(T) = \rho_0 + AT^{\alpha}$, where $\rho_0$ and $A$ are constants. With increasing $x$, the normal state $\rho$ decreases and $\alpha$ slightly increases. The $\alpha$ values (~1) and their trend (slightly increasing with increasing $x$) are similar to those of single crystals [22]. The $\alpha$ value of Ba122:P exhibits a striking deviation from standard Fermi-liquid theory where $\alpha$ = 2, which is a characteristic of Ba122:P [22]. $T_c$ exhibits dome-like behavior, which is consistent with the optimum $x$ = 0.3 reported for single crystals [22]. The $T_c$ values of the $x$ = 0.28 sample are the highest ($T_c^{\text{onset}}$ = 26.6 K and $T_c^{\text{zero}}$ = 22.4 K), although they are slightly lower than those of single crystals with the same $x$ [22]. This can be attributed to the strain effect, which is also observed in epitaxial Ba122:P films on single crystals [27, 28].

Figure 3 shows the transport $J_c$ plotted against $H$ for four Ba122:P films with $x$ = 0.22–0.42 grown at the optimum $T_s$ = 1050 °C. The $x$ = 0.22 and 0.28 films have high self-field $J_c$ at 4 K (~1 MA/cm$^2$), while those of the $x$ = 0.32 and 0.42 films (i.e., overdoped samples) are lower (~10$^5$ A/cm$^2$) than the $x$ = 0.22 and 0.28 ones, which is mainly because of their lower $T_c$ values. $J_c$ of the $x$ = 0.22 and 0.28 samples at 9 T is around 0.1 MA/cm$^2$ (see Fig. 3(b)). This value is roughly one order of magnitude lower than that at 9 T of an Fe(Se,Te) coated conductor [15]. The main reason would be the difference in the defect type working as pinning centers. It has been reported that the vertical defect in Ba122:P is dominant [17], while point-like defect is working as effective pinning in the Fe(Se,Te) conductor [15]. All of the $J_c$ values parallel to the *ab* plane are higher than those parallel to the *c* axis. This behavior is more remarkable at 12 K and under higher *H*, which is also observed in the films on IBAD-MgO with larger in-plane alignment of the top MgO layer of $\Delta\phi_{MgO}$ = 8° [17, 18]. This suggests that vortex pinning centers along the *c* axis act well at low temperatures, such as 4 K, but their performance decreases with increasing temperature. For the overdoped *x* = 0.42 film, its $J_c$ value under 9 T is as low as ≤1×10$^4$ A/cm$^2$ even at 4 K. From the viewpoint of the self-field and in-field $J_c$ as well as $T_c$ (see Fig. 2), we conclude that the *x* = 0.28 film exhibits the best performance among the films grown at the optimum $T_s$. We then investigated the vortex pinning properties of this film.

Figure 4 shows the $\theta_H$ dependence of $J_c$ at 4 and 12 K for the optimum *x* = 0.28 film grown at the optimum $T_s$ = 1050 °C. At 4 K, $J_c$ || *ab* is slightly higher than $J_c$ || *c* mainly because of its relatively low density of vertical defects (i.e., *c*-axis pinning centers) as observed in Ref. [17], i.e., the intrinsic electronic anisotropy of Ba122:P is dominant for this vortex pinning property. Relatively isotropic $J_c$ behavior is observed even at 9 T. The $J_c^{min}/J_c^{max}$ ratios at 4 K are 0.79, 0.75, 0.81, and 0.68 for 1, 3, 6, and 9 T, respectively. These values are comparable with those of the films on MgO single-crystals ($J_c^{min}/J_c^{max}$ ratios at 4 K of 0.94 (1 T), 0.86 (3T), 0.82 (6 T), and 0.77 (9 T)) [28]. However, the behaviour at 12 K is different from that of the films on MgO single-crystals, especially at 1 T (i.e., low magnetic field). The $J_c^{min}/J_c^{max}$ ratios at 12 K are 0.80, 0.82, 0.65, and 0.37 for 1, 3, 6, and 9 T, respectively, mainly because of much higher $J_c$ || *ab*, especially at 9 T. This suggests that vortex pinning centers along the *c* axis act well at low temperatures, but their effects decrease with increasing temperature. This may be an origin of a clear $J_c$ peak || *ab* especially observed at 9T and 12 K. In the case of the films on MgO single-crystals [28], the $J_c^{min}/J_c^{max}$ ratio at 12 K and 1 T is

0.79, which is almost the same value as that of the film on IBAD-MgO. However, a $J_c$ peak $\parallel c$ was clearly observed in the case of single-crystal sample mainly because of vertical defects effectively acting as $c$-axis pinning centers. These isotropic $J_c$ properties are useful for fabrication of Ba122:P coated conductors.

From the viewpoint of the crystallinity, $T_c$, and $J_c$, we conclude that the optimum $T_s$ and $x$ in the film are 1050 °C and $x = 0.28$, respectively. However, we found an unexpected result that although the sample grown at a higher of $T_s = 1250$ °C exhibits a broad $\Delta\omega$ (FWHM ≈ 3°) and has lower crystallinity (see Fig. 1(c)), its $T_c$ is comparable with that of the optimized film ($T_c^{onset} = 26.6$ and $T_c^{zero} = 23.8$ K, data not shown) and it exhibits unexpectedly higher $J_c$ under an external magnetic field, especially parallel to the $ab$ plane direction, than those of the optimized samples. Here, we would like to comment that we did not measure $J_c$ properties of the film grown at $T_s = 875$ °C exhibiting a similar broad $\Delta\omega$ (FWHM ≈ 3°, see Fig. 1(c)). Because a weak-link behavior originating from granularity [33] due to the low temperature growth for Sr122:Co (700 °C [34]) was clearly observed, irrespective of its epitaxial growth. From this result, we speculated that the $J_c$ of such low-temperature growth films is also very low due to the weak link.

For this high-$T_s$ growth film ($T_s = 1250$ °C), we used a nominal $x = 0.4$ PLD target. Figure 5 shows $J_c$ plotted against $H$ for the film grown at $T_s = 1250$ °C. The values of the sample grown at the optimum $T_s$ and $x$, taken from the upper right panel of Fig. 3(a), are shown for comparison. As shown in Fig. 5(a), the $J_c \parallel c$ values are comparable, while the $J_c \parallel ab$ values under higher $H$ are much higher than those of the film with the optimized values. This trend is confirmed also by the clear $J_c$ peaks $\parallel ab$ in Fig. 5(b) for all $H$ values. The intrinsic electronic anisotropy of Ba122:P cannot simply explain the clear $J_c$ peaks ∥ $ab$ (i.e., enhancement of pinning parallel to $ab$). One of the possible origins is the increase in the concentration of horizontal defects, such as stacking faults, along the $ab$ plane, which are observed in a Ba122:Co film grown by low-power KrF excimer laser PLD [35]. This speculation is supported by the results of the out-of-plane rocking curve measurements in Fig. 1(c). High-$T_s$ growth of 1250 °C leads to a broad FWHM of $\Delta\omega$ of >3°, suggesting increasing density of horizontal defects.

4. Summary

We optimized $T_s$ and $x$ for $BaFe_2(As_{1-x}P_x)_2$ (Ba122:P) thin films on practical IBAD-MgO metal-tape substrates. We found that the optimum $T_s$ and $x$ values are 1050 °C and $x = 0.28$, respectively, from the viewpoint of the crystallinity and $T_c$. The optimized films exhibit high $T_c^{onset} = 26.6$ and $T_c^{zero} = 22.4$ K along with high self-field $J_c$ at 4 K (~1 MA/cm$^2$) and isotropic $J_c$ under an external magnetic field. Additionally, we found that lower crystallinity samples grown at $T_s = 1250$ °C exhibit higher $J_c$ under an external magnetic field, especially parallel to the $ab$ plane, than the optimized samples. Such high $J_c$ performance would originate from horizontal defects, such as stacking faults, acting as strong vortex pinning centers. This result indicates that as well as the crystallinity and doping concentration, introducing an optimum concentration of defects is also important to achieve strong vortex pinning properties for Ba122:P films; a similar effect is known also for single crystal because vortex pinning in pure and high quality single crystals is rather poorer than that in poorer quality ones. The critical current of the Ba122:P coated conductors is still to be improved by increasing their thickness to achieve a critical current comparable to cuprate coated conductors.

Acknowledgments

This work was supported by the Ministry of Education, Culture, Sports, Science and Technology (MEXT) through the Element Strategy Initiative to Form Core Research Center. H. Hi. was also supported by the Japan Society for the Promotion of Science (JSPS) through a Grant-in-Aid for Young Scientists (A) (Grant Number 25709058), a JSPS Grant-in-Aid for Scientific Research on Innovative Areas "Nano Informatics" (Grant Number 25106007), and Support for Tokyotech Advanced Research (STAR).

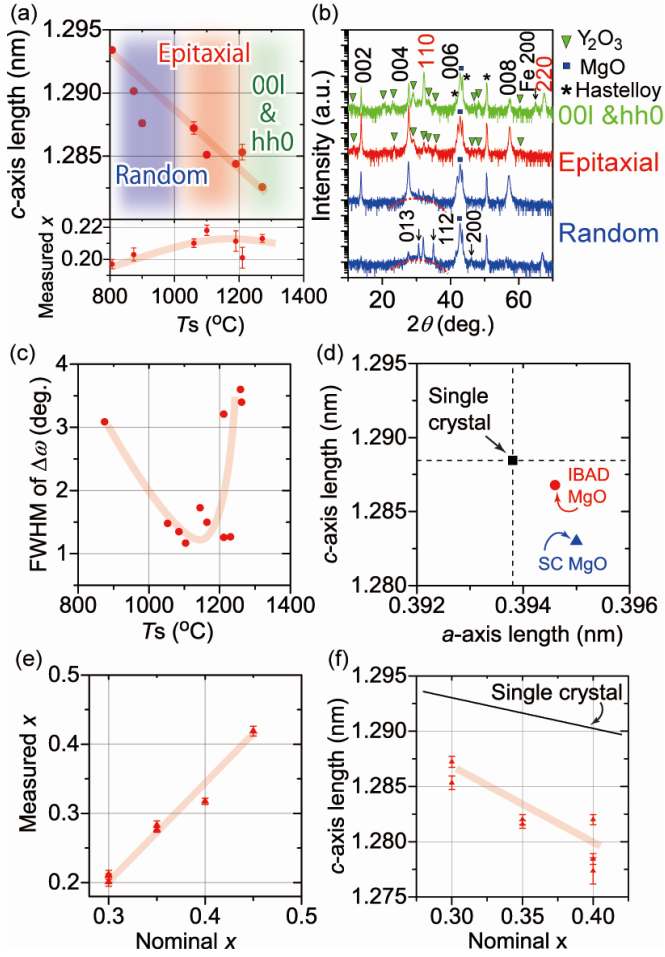

**Figure 1.** Film structure and chemical composition of $BaFe_2(As_{1-x}P_x)_2$ ($x$ = 0.22–0.42) films. (a) Substrate temperature ($T_s$) dependence of the $c$-axis length and phosphorous concentration [$x$ = P/(As+P))] in films grown using a $BaFe_2(As_{0.7}P_{0.3})_2$ (i.e., $x$ = 0.3) PLD target. Three regions with different orientations ('Random', 'Epitaxial', and '00$l$ & $hh$0' mixture) are shown. (b) Typical out-of-plane XRD patterns of films grown in the three regions in (a). From bottom to top, $T_s$ increases from 800 to 1250 °C. The two red dotted curves around $2\theta \approx 30°$ for the randomly oriented samples indicate the halo patterns from the amorphous $Y_2O_3$ bed-layer in the IBAD-MgO substrate. At $T_s \geq$ 1000 °C, the amorphous $Y_2O_3$ bed-layer crystallizes, as indicated by the inverted triangles. (c) FWHM of the out-of-plane rocking curves of the 004 diffraction of the films in (a) as a function of $T_s$. (d) Relationship between the *c*- and *a*-axis lengths of the film grown at the optimum $T_s$ = 1050 °C using a $x$ = 0.3 PLD target (IBAD MgO). The values for a single crystal [21] and an epitaxial film on a MgO single-crystal (SC MgO) [27] with the same $x$ are shown for comparison. (e) Relationship between $x$ in films grown at the optimum $T_s$ = 1050 °C and the nominal $x$ in the PLD targets. (f) Relationship between the *c*-axis length of films grown at the optimum $T_s$ = 1050 °C and the nominal $x$ in the PLD targets. The relationship for films grown on single crystals is shown for comparison (thin black line) [21].

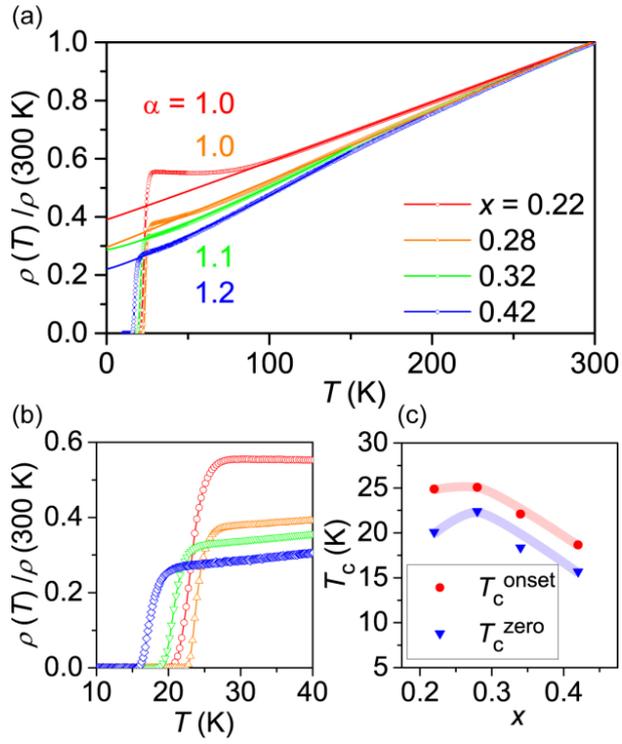

**Figure 2.** (a, b) Temperature ($T$) dependence of normalized resistivity ($\rho$) for BaFe$_2$(As$_{1-x}$P$_x$)$_2$ ($x$ = 0.22–0.42) films grown at the optimum $T_s$ = 1050 °C. The straight lines and $\alpha$ values in (a) are the fitting results using the power law, $\rho(T) = \rho_0 + AT^\alpha$, where $\rho_0$ and $A$ are constants. (b) Enlarged image around $T_c$ in (a). (c) $x$ dependence of $T_c$.

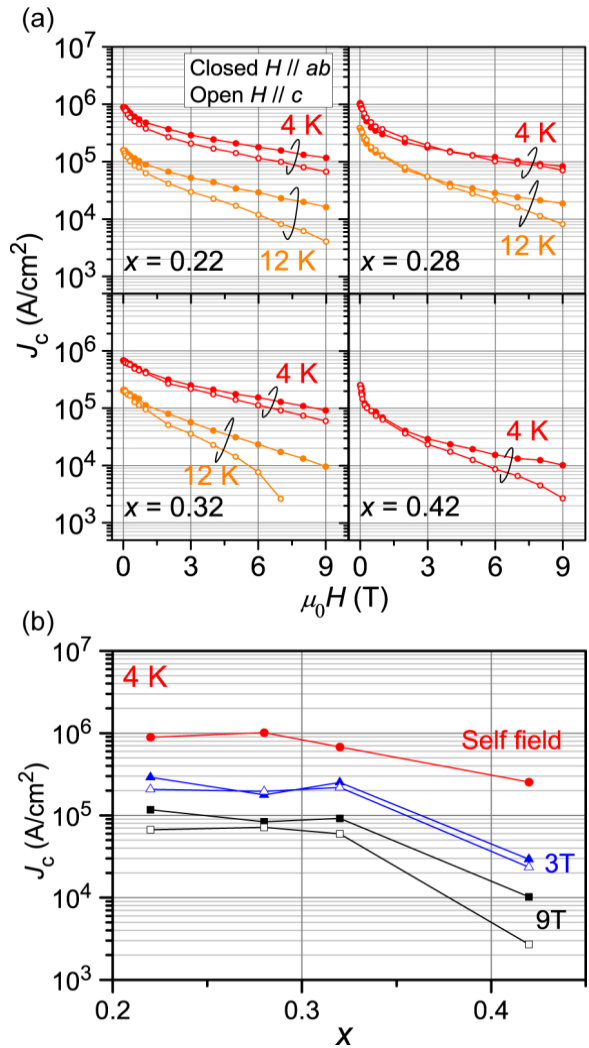

**Figure 3.** External magnetic field ($H$) dependence of $J_c$ for BaFe$_2$(As$_{1-x}$P$_x$)$_2$ ($x$ = 0.22–0.42) films grown at the optimum $T_s$ = 1050 °C. The closed and open symbols indicate the configurations of $H \parallel ab$ and $H \parallel c$ of the films, respectively. (a) Each measurement temperature and chemical composition $x$ in the films are shown in each figure. (b) $x$ dependence of $J_c$ at 4 K under the self field (red), 3 T (blue), and 9 T (black). All the data are taken from (a).

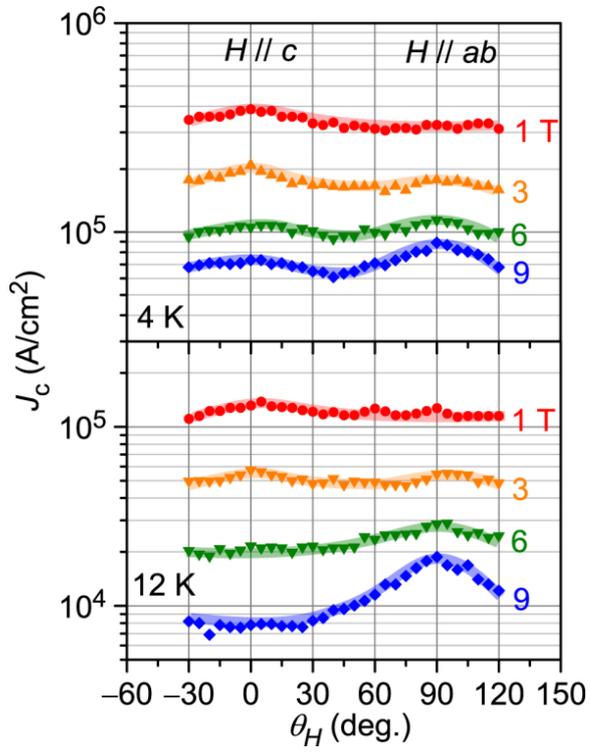

**Figure 4.** External magnetic field angle ($\theta_H$) dependence of $J_c$ at 4 K (top) and 12 K (bottom) for the BaFe$_2$(As$_{0.72}$P$_{0.28}$)$_2$ film grown at the optimum $T_s$ = 1050 °C. $\theta_H$ = 0 and 90° indicate that $H$ is parallel to the $c$ axis and $ab$ plane of the film, respectively.

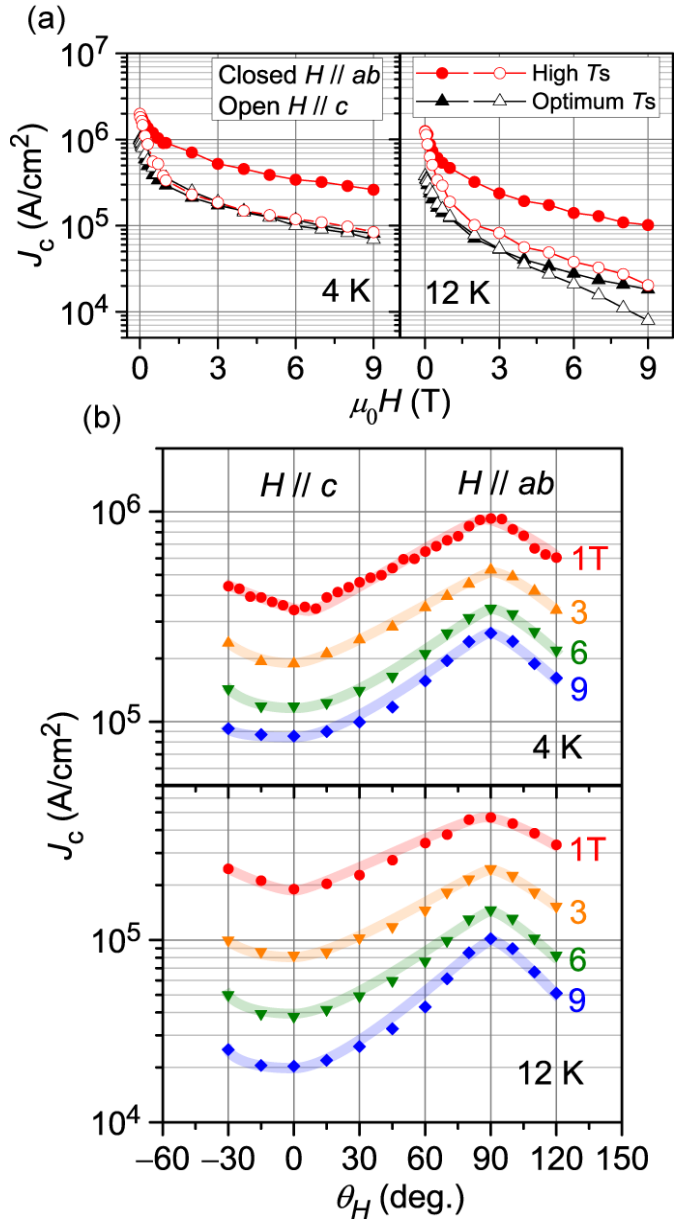

**Figure 5.** (a) $H$ and (b) $\theta_H$ dependence of $J_c$ at 4 K and 12 K for the BaFe$_2$(As$_{0.66}$P$_{0.34}$)$_2$ film grown at $T_s$ = 1250 °C. The results for the BaFe$_2$(As$_{0.72}$P$_{0.28}$)$_2$ film grown at the optimum $T_s$ = 1050 °C are shown in (a) for comparison.